\documentstyle{article}

\begin{document}

\title{On the nature of the force acting on a charged classical particle deviated
from its geodesic path in a gravitational field}
\author{Vesselin Petkov \\
Science College, Concordia University\\
1455 De Maisonneuve Boulevard West\\
Montreal, Quebec, Canada H3G 1M8\\
E-mail: vpetkov@alcor.concordia.ca}
\date{23 October 2001}
\maketitle

\begin{abstract}
In general relativity the gravitational field is a manifestation of
spacetime curvature and unlike the electromagnetic field is not a force
field. A particle falling in a gravitational field is represented by a
geodesic worldline which means that no force is acting on it. If the
particle is at rest in a gravitational field, however, its worldline is no
longer geodesic and it is subjected to a force. The nature of that force is
an open question in general relativity. The aim of this paper is to outline
an approach toward resolving it in the case of classical charged particles
which was initiated by Fermi in 1921.
\end{abstract}

General relativity provides a consistent no-force explanation of
gravitational interaction of bodies following geodesic paths. However, it is
silent on the nature of the very force we regard as gravitational - the
force acting upon a body deviated from its geodesic path due to its being at
rest in a gravitational field.

In both special relativity (in flat spacetime) and general relativity (in
curved spacetime) a particle offering no resistance to its motion is
represented by a geodesic worldline. As the non-resistant motion of a
particle is regarded as inertial a particle whose worldline is geodesic is
moving by inertia. In both special and general relativity a particle whose
worldline is not geodesic is prevented from moving by inertia and therefore
is subjected to an inertial force. Hence a particle supported in a
gravitational field is deviated from its geodesic path (i.e. prevented from
moving by inertia) which means that the force acting on it is not
gravitational but inertial in origin.

The mass causing the spacetime curvature determines the shape of the
particle's geodesic worldline, and in general which reference frames are
inertial \cite{weinberg}, but the force arising when the particle is
deviated from its geodesic path originates neither from that mass nor from
the distant masses (as Mach proposed). This force has the same origin as the
force acting on a test particle prevented from following a geodesic path in
an empty spacetime. It should be stressed that in general relativity the
force acting on a particle deviated from its geodesic path due to its being
at rest in a gravitational field is non-gravitational in origin. As Rindler
put it ''ironically, instead of explaining inertial forces as
gravitational... in the spirit of Mach, Einstein explained gravitational
forces as inertial'' \cite{rindler}. This is the reason why ''there is no
such thing as the force of gravity'' in general relativity \cite{synge}.

Here it will be shown that a corollary of general relativity - that the
propagation of light in a gravitational field is anisotropic - in
conjunction with the classical electromagnetic mass theory \cite{thomson}-
\cite{rohrlich} sheds some light on the nature of the force acting on a
classical charged particle deviated from its geodesic path.

Consider a classical electron \cite{classical} at rest in the non-inertial
reference frame $N^{g}$ of an observer supported in the Earth's
gravitational field. Following Lorentz \cite{lorentz} and Abraham \cite
{abraham} we assume that the electron charge is uniformly distributed on a
spherical shell. The repulsion of the charge elements of an electron in
uniform motion in flat spacetime cancels out exactly and there is no net
force acting on the electron. As we shall see below, however, the average
anisotropic velocity of light in $N^{g}$ (i) gives rise to a self-force
acting on an electron deviated from its geodesic path by disturbing the
balance of the mutual repulsion of its charge elements, and (ii) makes a
free electron fall in $N^{g}$ with an acceleration ${\bf g}$ in order to
balance the repulsion of its charge elements. No force is acting upon a
falling electron (whose worldline is geodesic) but if it is prevented from
falling (i.e. deviated from its geodesic path) the average velocity of light
with respect to it becomes anisotropic and disturbs the balance of the
mutual repulsion of the elements of its charge which results in a self-force
trying to force the electron to fall. This force turns out to be equal to
the gravitational force ${\bf F}=m^{g}{\bf g}$, where $m^{g}=U/c^{2}$
represents the passive gravitational mass of the classical electron and $U$
is the energy of its field. As the coefficient $m^{g}$ in front of ${\bf g}$
is exactly equal to $U/c^{2}$ (without the $4/3$ factor) it turns out that
the mass of the {\em classical} electron is purely electromagnetic in origin
when the average anisotropic velocity of light in a gravitational field is
taken into account.

In 1921 Fermi \cite{fermi} studied the nature of the force acting on a
charge at rest in a gravitational field of strength $g$ in the framework of
general relativity and the classical electromagnetic mass theory. The
potential
\begin{equation}
\varphi =\frac{e}{4\pi \epsilon _{0}r}\left( 1-\frac{1}{2}\frac{gz}{c^{2}}%
\right)  \label{1fermi}
\end{equation}
he derived, however, contains the $1/2$ factor in the parenthesis which
leads to a contradiction with the principle of equivalence when the electric
field is calculated from this potential: it follows from (\ref{1fermi}) that
the electric field of a charge supported in the Earth's gravitational field
coincides with the instantaneous electric field of a charge moving with an
acceleration ${\bf a}=-{\bf g}/2$ (obviously the principle of equivalence
requires that ${\bf a}=-{\bf g}$). Now we shall show that the average
anisotropic velocity of light in a gravitational field gives rise to a
Li\'{e}nard-Wiechert-like contribution to the potential $\varphi $ which
removes the $1/2$ factor in (\ref{1fermi}).

Why the average velocity of light between two points in a gravitational
field is not equal to $c$ can be most clearly shown by considering two extra
light rays parallel and anti-parallel to the gravitational acceleration $%
{\bf g}$ in the Einstein thought experiment involving an elevator at rest in
the Earth's gravitational field (see Figure~1) \cite{petkov}.

\begin{center}
\begin {picture}(0,100)(-20,130)
\setlength{\unitlength}{0.25mm}
\put(-100,100){\framebox(110,220){}}
\put(10,212){\circle*{4}}
\put(-6,217){$B$}
\put(29,212){\line(1,0){10}}
\put(34,201){\vector(0,1){11}}
\put(10,190){\circle*{4}}
\put(-6,175){$B^{\prime}$}
\put(29,190){\line(1,0){10}}
\put(34,201){\vector(0,-1){11}}
\put(40,198){$\delta =\frac{1}{2} gt^{2}=gr^{2}/2c^{2}$}
\put(-5,305){$A$}
\put(10,320){\circle*{4}}
\put(10,100){\circle*{4}}
\put(-5,105){$C$}
\put(-100.5,212){\circle*{4}}
\put(-95,217){$D$}

\qbezier(-100.5,212)(-20,212)(2.5,193)
\put(2.5,193.3){\vector(3,-2){2}}
\put(-121,96.6){---}
\put(-121,317.1){---}
\put(-114,200){\vector(0,1){119.5}}
\put(-114,220.1){\vector(0,-1){119}}
\put(-131,212){$2r$}
\put(-100,80){\line(0,1){10}}
\put(10,80){\line(0,1){10}}
\put(-45,85){\vector(1,0){53.7}}
\put(-45,85){\vector(-1,0){53.7}}
\put(-50,87.5){$r$}
\put(17,320.5){\vector(0,-1){129}}
\put(17,99){\vector(0,1){89}}
\thicklines
\put(-45,281){\vector(0,-1){33}}
\put(-39,267){{\bf g}}
\end {picture}
\end{center}

\vspace{2cm}

\begin{center}
\begin{list}{}{\leftmargin=1.5em \rightmargin=0em}\item[]
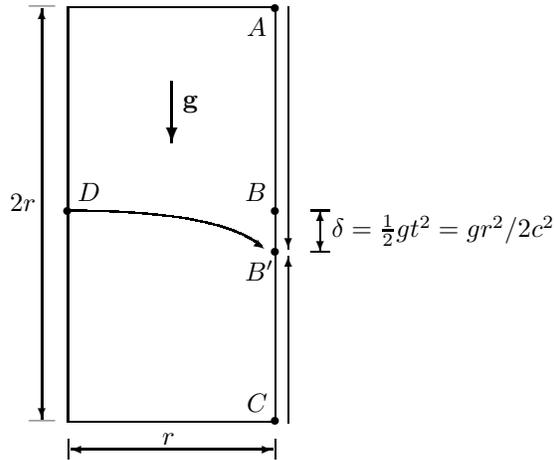
{\bf Figure 1}. Three light rays propagate in an elevator at rest
in the Earth's gravitational field. After having been emitted
simultaneously from points $A$, $C$, and $D$ the rays meet at $B^{\prime}$
(the ray propagating from $D$ toward $B$, but arriving at $B^{\prime}$,
represents the original thought experiment considered by
Einstein). The light rays emitted from $A$ and $C$ are introduced
in order to determine the expression for the average
velocity of light in a gravitational field. It takes the same
coordinate time $t=r/c$ for the rays to travel the distances $DB^{\prime}
\approx r$, $AB^{\prime}=r+\delta$, and $CB^{\prime}=r-\delta$. Therefore the
average velocity of the downward ray from $A$ to $B^{\prime}$ is
${c}_{AB^{\prime}}= (r+\delta)/t \approx c(1+gr/2c^{2})$; the
average velocity of the upward ray from $C$ to $B^{\prime}$ is
${c}_{CB^{\prime}}= (r-\delta)/t \approx c(1-gr/2c^{2}).$
\end{list}
\end{center}

Three light rays are emitted simultaneously in the elevator (representing a
non-inertial reference frame $N^{g}$) from points $A$, $C$, and $D$ toward
point $B$. The emission of the rays is also simultaneous in a reference
frame $I$ (a local Lorentz frame) which is momentarily at rest with respect
to $N^{g}$. At the moment the light rays are emitted $I$ starts to fall in
the gravitational field. At the next moment an observer in $I$ sees that the
elevator moves upward with an acceleration $g=\left| {\bf g}\right| $.
Therefore as seen from $I$ the three light rays arrive simultaneously not at
point $B$, but at $B^{\prime }$ since for the time $t=r/c$ the elevator
moves at a distance $\delta =gt^{2}/2=gr^{2}/2c^{2}$. As the simultaneous
arrival of the three rays at the point $B^{\prime }$ in $I$ is an absolute
event (the same in all reference frames) being a {\em point} event, it
follows that the rays arrive simultaneously at $B^{\prime }$ as seen from $%
N^{g}$ as well. Since for the {\em same} coordinate time $t=r/c$ in $N^{g}$
the three light rays travel different distances $DB^{\prime }\approx r$, $%
AB^{\prime }=r+\delta $, and $CB^{\prime }=r-\delta $ before arriving
simultaneously at point $B^{\prime }$ an observer in the elevator concludes
that the {\em average} velocity of the light ray propagating from $A$ to $%
B^{\prime }$ is slightly greater than $c$

\[
{c}_{AB^{\prime}}^{g}=\frac{r+\delta}{t} \approx c\left( 1+\frac{gr}{2c^{2}}%
\right).
\]

\noindent The average velocity ${c}_{CB^{\prime}}^{g}$ of the light ray
propagating from $C$ to $B^{\prime}$ is slightly smaller than $c$

\[
{c}_{CB^{\prime }}^{g}=\frac{r-\delta }{t}\approx c\left( 1-\frac{gr}{2c^{2}}%
\right) .
\]
It is easily seen that to within terms $\sim c^{-2}$ the average light
velocity between $A$ and $B$ is equal to that between $A$ and $B^{\prime }$,
i.e. $c_{AB}^{g}=c_{AB^{\prime }}^{g}$ and also $c_{CB}^{g}=c_{CB^{\prime
}}^{g}$:
\begin{equation}
c_{AB}^{g}=\frac{r}{t-\delta /c}\approx c\left( 1+\frac{gr}{2c^{2}}\right)
\label{c_g-AB}
\end{equation}
and
\begin{equation}
c_{CB}^{g}=\frac{r}{t+\delta /c}\approx c\left( 1-\frac{gr}{2c^{2}}\right) .
\label{c_g-CB}
\end{equation}
As the average velocities (\ref{c_g-AB}) and (\ref{c_g-CB}) are not
determined with respect to a specific point and since the {\em coordinate}
time $t$ is involved in their calculation, it is clear that the expressions (%
\ref{c_g-AB}) and (\ref{c_g-CB}) represent the average {\em coordinate}
velocities between the points $A$ and $B$ and $C$ and $B$, respectively.

These expressions for the average coordinate velocity of light in $N^{g}$
can be also obtained from the coordinate velocity of light at a point in a
{\em parallel} gravitational field. If the $z$-axis is antiparallel to the
elevator's acceleration ${\bf g}$ the spacetime metric in $N^{g}$ has the
form \cite{misner}
\begin{equation}
ds^{2}=\left( 1+\frac{2gz}{c^{2}}\right) c^{2}dt^{2}-dx^{2}-dy^{2}-dz^{2}
\label{ds_g}
\end{equation}
from where the coordinate velocity of light at a point $z$ in a parallel
gravitational field is immediately obtained (for $ds^{2}=0$)
\begin{equation}
c^{g}\left( z\right) =c\left( 1+\frac{gz}{c^{2}}\right) .  \label{c_g_coord}
\end{equation}
Notice that (\ref{ds_g}) is the standard spacetime interval in a {\em %
parallel} gravitational field \cite{misner}, which does not coincide with
the expression for the spacetime interval in a spherically symmetric
gravitational field (i.e. the Schwarzschild metric) \cite[p. 395]{ohanian}
\begin{equation}
ds^{2}=\left( 1-\frac{2GM}{c^{2}r}\right) c^{2}dt^{2}-\left( 1+\frac{2GM}{%
c^{2}r}\right) \left( dx^{2}+dy^{2}+dz^{2}\right) .  \label{schwarz}
\end{equation}
The metric (\ref{ds_g}) can be written in a form similar to (\ref{schwarz})
if we choose $r=r_{0}+z$ where $r_{0}$ is a constant

\begin{equation}
ds^{2}=\left( 1-\frac{2GM}{c^{2}\left( r_{0}+z\right) }\right)
c^{2}dt^{2}-\left( dx^{2}+dy^{2}+dz^{2}\right) .  \label{ds_g1}
\end{equation}
As $g=GM/r_{0}^{2}$ and for $z/r_{0}<1$ we can write

\begin{equation}
ds^{2}=\left( 1-\frac{2GM}{c^{2}r_{0}}+\frac{2gz}{c^{2}}\right)
c^{2}dt^{2}-\left( dx^{2}+dy^{2}+dz^{2}\right) .  \label{ds_g2}
\end{equation}
As the gravitational potential is undetermined to within an additive
constant we can choose $GM/r_{0}=0$ in (\ref{ds_g2}); more precisely, when
calculating the gravitational potential we can set the constant of
integration to be equal to $-GM/r_{0}.$ With this choice of the integration
constant (\ref{ds_g2}) coincides with (\ref{ds_g}). Although similar (\ref
{ds_g1}) and (\ref{schwarz}) have different values for $g_{ii}$ $(i=1,2,3)$:
$g_{ii}=-1$ in (\ref{ds_g1}), whereas $g_{ii}=-\left( 1+2GM/c^{2}r\right) $
in (\ref{schwarz}). This reflects the fact that in a parallel gravitational
field proper and coordinate times do not coincide (except for the proper
time of an observer at infinity) whereas proper and coordinate distances are
the same \cite{rindler68}.

Throughout the paper we will be concerned only with a parallel gravitational
field. As all effects we will be studying involve distances of the order of
the classical electron radius ($\sim 10^{-15}$ $m$) that approximation is an
excellent one. It is also clear that the acceleration $g$ will have the same
value in all points of a region of such dimensions.

Using the coordinate velocity (\ref{c_g_coord}) we obtain for the average
coordinate velocity of light propagating between $A$ and $B$ (Figure 1)
\[
c_{AB}^{g}=\frac{1}{2}\left( c_{A}^{g}+c_{B}^{g}\right) =\frac{1}{2}\left[
c\left( 1+\frac{gz_{A}}{c^{2}}\right) +c\left( 1+\frac{gz_{B}}{c^{2}}\right) %
\right]
\]
and as $z_{A}=z_{B}+r$
\begin{equation}
c_{AB}^{g}=c\left( 1+\frac{gz_{B}}{c^{2}}+\frac{gr}{2c^{2}}\right) .
\label{c_g__AB}
\end{equation}
For the average coordinate velocity of light propagating between $B$ and $C$
we obtain
\begin{equation}
c_{BC}^{g}=c\left( 1+\frac{gz_{B}}{c^{2}}-\frac{gr}{2c^{2}}\right)
\label{c_g_CB}
\end{equation}
since $z_{C}=z_{B}-r$. When the coordinate origin is at point $B$ ($z_{B}=0$%
) the expressions (\ref{c_g__AB}) and (\ref{c_g_CB}) coincide with (\ref
{c_g-AB}) and (\ref{c_g-CB}).

There exists a third way to derive the average coordinate velocity of light
in $N^{g}$. As the coordinate velocity $c^{g}\left( z\right) $ (\ref
{c_g_coord}) is continuous on the interval $\left[ z_{A},z_{B}\right] $ in
the case of weak parallel gravitational fields one can calculate the average
coordinate velocity between $A$ and $B$:

\[
c_{AB}^{g}=\frac{1}{z_{B}-z_{A}}\int_{z_{A}}^{z_{B}}c^{g}\left( z\right)
dz=c\left( 1+\frac{gz_{B}}{c^{2}}+\frac{gr}{2c^{2}}\right) .
\]
As expected this expression coincides with (\ref{c_g__AB}).

The average coordinate velocities (\ref{c_g__AB}) and (\ref{c_g_CB})
correctly describe the propagation of light in $N^{g}$ yielding the right
expression $\delta =gr/2c^{2}$ (see Figure 1). It should be stressed that
without these average coordinate velocities the fact that the light rays
emitted from $A$ and $C$ arrive not at $B,$ but at $B^{\prime }$ cannot be
explained.

As a coordinate velocity, the average coordinate velocity of light is not
determined with respect to a specific point and depends on the choice of the
coordinate origin. Also, it is the same for light propagating from $A$ to $B$
and for light travelling in the opposite direction, i.e. $%
c_{AB}^{g}=c_{BA}^{g}$. Therefore, like the coordinate velocity (\ref
{c_g_coord}) the average coordinate velocity is also isotropic. Notice,
however, that the average coordinate velocity of light is isotropic in a
sense that the average light velocity between two points is the same in both
directions. But as seen from (\ref{c_g__AB}) and (\ref{c_g_CB}) the average
coordinate velocity of light between different pairs of points, whose points
are the same distance apart, is different. As a result as seen in Figure 1
the light ray emitted at $A$ arrives at $B$ before the light ray emitted at $%
C$.

The average coordinate velocity of light explains the propagation of light
in the Einstein elevator, but cannot be used in a situation where the
average light velocity between two points (say a source and an observation
point) is determined with respect to one of the points. Such situations
occur, as we shall see, when one calculates the potential, the electric
field, and the self-force of a charge in a gravitational field. As the local
velocity of light is $c$ the average velocity of light between a source and
an observation point depends on which of the two points is regarded as a
reference point with respect to which the average velocity is determined (at
the reference point the local velocity of light is $c$). The dependence of
the average velocity on which point is chosen as a reference point
demonstrates that that velocity is anisotropic. This anisotropic velocity
should be regarded as an average{\em \ proper} velocity of light since it is
determined with respect to a given point and its calculation involves the
proper time at that point.

Consider a light source at point $B$ (Figure 1). To calculate the average
proper velocity of light originating from $B$ and observed at $A$ (that is,
as seen from $A$) we have to determine the initial velocity of a light
signal at $B$ and its final velocity at $A$ both with respect to $A$. As the
local velocity of light is $c$ the final velocity of the light signal
determined at $A$ is also $c$. Its initial velocity at $B$ as seen from $A$
is
\[
c_{B}^{g}=\frac{dz_{B}}{d\tau _{A}}=\frac{dz_{B}}{dt}\frac{dt}{d\tau _{A}}
\]
where $dz_{B}/dt=c^{g}\left( z_{B}\right) $ is the coordinate velocity (\ref
{c_g_coord}) at $B$%
\[
c^{g}\left( z_{B}\right) =c\left( 1+\frac{gz_{B}}{c^{2}}\right)
\]
and $d\tau _{A}$ is the proper time at $A$%
\[
d\tau _{A}=\left( 1+\frac{gz_{A}}{c^{2}}\right) dt.
\]
Since $z_{A}=z_{B}+r$ for the coordinate time $dt$ we have
\[
dt=\left( 1-\frac{gz_{A}}{c^{2}}\right) d\tau _{A}=\left( 1-\frac{gz_{B}}{%
c^{2}}-\frac{gr}{c^{2}}\right) d\tau _{A}.
\]
Then for the initial velocity $c_{B}^{g}$ at $B$ as seen from $A$ we obtain
\[
c_{B}^{g}=c\left( 1+\frac{gz_{B}}{c^{2}}\right) \left( 1-\frac{gz_{B}}{c^{2}}%
-\frac{gr}{c^{2}}\right)
\]
or keeping only the terms $\sim c^{-2}$
\[
c_{B}^{g}=c\left( 1-\frac{gr}{c^{2}}\right) .
\]
For the {\em average proper} velocity $c_{BA}^{g}=(1/2)(c_{B}^{g}+c)$ of
light propagating from $B$ to $A$ {\em as seen from} $A$ we have
\begin{equation}
c_{BA}^{g}\left( as\ seen\ from\ A\right) =c\left( 1-\frac{gr}{2c^{2}}%
\right) .  \label{c-g_BA}
\end{equation}

As the local velocity of light at $A$ is $c$ it follows that if light
propagates from $A$ toward $B$ its average proper velocity $c_{AB}^{g}\left(
as\ seen\ from\ A\right) $ will be equal to the average proper velocity of
light propagating from $B$ toward $A$ $c_{BA}^{g}\left( as\ seen\ from\
A\right) $. Thus, as seen from $A$, the back and forth average proper
velocities of light travelling between $A$ and $B$ are the {\em same}.

Now let us determine the average proper velocity of light between $B$ and $A$
with respect to the source point $B$. A light signal emitted at $B$ as seen
from $B$ will have an initial (local) velocity $c$ there. The final velocity
of the signal at $A$ as seen from $B$ will be
\[
c_{A}^{g}=\frac{dz_{A}}{d\tau _{B}}=\frac{dz_{A}}{dt}\frac{dt}{d\tau _{B}}
\]
where $dz_{A}/dt=c^{g}\left( z_{A}\right) $ is the coordinate velocity at $A$%
\[
c^{g}\left( z_{A}\right) =c\left( 1+\frac{gz_{A}}{c^{2}}\right)
\]
and $d\tau _{B}$ is the proper time at $B$%
\[
d\tau _{B}=\left( 1+\frac{gz_{B}}{c^{2}}\right) dt.
\]
Then as $z_{A}=z_{B}+r$ for $c_{A}^{g}$ we obtain
\[
c_{A}^{g}=c\left( 1+\frac{gr}{c^{2}}\right)
\]
and the average proper velocity of light propagating from $B$ to $A$ {\em as
seen from }$B$ becomes
\begin{equation}
c_{BA}^{g}\left( as\ seen\ from\ B\right) =c\left( 1+\frac{gr}{2c^{2}}%
\right) .  \label{c-g-AB}
\end{equation}
If a light signal propagates from $A$ to $B$ its average proper velocity $%
c_{AB}^{g}\left( as\ seen\ from\ B\right) $ will be equal to that from $B$
to $A$ $c_{BA}^{g}\left( as\ seen\ from\ B\right) $. Comparing (\ref{c-g_BA}%
) and (\ref{c-g-AB}) demonstrates that the two average proper velocities
between the same points are not equal and depend on from where they are
seen. As we expected the fact that the local velocity of light at the
reference point is $c$ makes the average proper velocity between two points
dependant on where the reference point is.

In order to express the average proper velocity of light in a vector form
let the light emitted from $B$ be observed at different points. The average
proper velocity of light emitted at $B$ and determined at $A$ is given by (%
\ref{c-g_BA}). As seen from point $C$ the average proper velocity of light
from $B$ to $C$ will be given by an expression derived in the same way as (%
\ref{c-g-AB})
\[
c_{BC}^{g}\left( as\ seen\ from\ C\right) =c\left( 1+\frac{gr}{2c^{2}}%
\right) .
\]
As seen from a point $P$ at a distance $r$ from $B$ and lying on a line
forming an angle $\theta $ with the acceleration ${\bf g}$ the average
proper velocity of light from $B$ is
\[
c_{BP}^{g}\left( as\ seen\ from\ P\right) =c\left( 1+\frac{gr\cos \theta }{%
2c^{2}}\right) .
\]
Then the average proper velocity of light coming from $B$ as seen from a
point defined by the position vector ${\bf r}$ originating from $B$ has the
form
\begin{equation}
\bar{c}^{g}=c\left( 1+\frac{{\bf g}\cdot {\bf r}}{2c^{2}}\right) .
\label{c_g}
\end{equation}

As evident from (\ref{c_g}) the average proper velocity of light emitted
from a common source and determined at different points around the source is
anisotropic in $N^{g}$ - if the observation point is above the light source
the average proper velocity of light is slightly smaller than $c$\ and
smaller than the average proper velocity as determined from an observation
point below the source. If an observer at point $B$ (Figure 1) determines
the average proper velocities of light coming from $A$ and $C$ he finds that
they are also anisotropic - the average proper velocity of light coming from
$A$ is greater than that emitted at $C$. However, if the observer at $B$
(Figure 1) determines the back and forth average proper velocities of light
propagating between $A$ and $B$ (or between $B$ and $C$) he finds that they
are the same.

One deduces from (\ref{c_g}) that $|{\bf g}\cdot {\bf r}/2c^{2}|<1$ in order
that $\bar{c}^{g}$ be positive. In fact, that restriction is always
satisfied in all cases involving the principle of equivalence since it is
weaker than the one imposed by that principle which requires that only small
regions in a gravitational field where the field is parallel are considered
\cite{pe}. In the case of light travelling a large distance $h$ between two
points $A$ and $B$ ($r_{A}>r_{B}$) along the radial direction in a
gravitational field it can be shown that the average proper velocity of
light can be expressed in terms of the gravitational potentials of the
source and observation points. When propagating along the radial direction a
light signal does not ''feel'' the spacetime curvature and the coordinate
velocity (\ref{c_g_coord}) is used in the calculation of the average proper
velocity. For instance, the average proper velocity
\[
c_{BA}^{g}\left( as\ seen\ from\ A\right) =c\left( 1-\frac{gh}{2c^{2}}%
\right)
\]
of light propagating from $B$ to $A$ as seen from $A$ (which means that the
local light velocity at $A$ is $c$) can be written as
\[
c_{BA}^{g}\left( as\ seen\ from\ A\right) =c\left( 1+\frac{GM}{2c^{2}r_{A}}-%
\frac{GM}{2c^{2}r_{B}}\right) .
\]

The velocity (\ref{c_g}) demonstrates that there exists a directional
dependence in the propagation of light between two points in a non-inertial
frame of reference $N^{g}$ at rest in a gravitational field. This anisotropy
in the propagation of light has been an overlooked corollary of general
relativity. In fact, up to now neither the average coordinate velocity nor
the average proper velocity of light have been defined. However, we have
seen that the average coordinate velocity is needed to account for the
propagation of light in a gravitational field (to explain the fact that two
light signals emitted from points $A$, and $C$ in Figure 1 meet at $%
B^{\prime }$, not at $B$). We will also see below that the average proper
velocity of light is necessary for the correct description of
electromagnetic phenomena in a gravitational field.

The anisotropic velocity of light (\ref{c_g}) leads to two changes in the
scalar potential
\begin{equation}
d\varphi ^{g}=\frac{1}{4\pi \epsilon _{0}}\frac{\rho dV^{g}}{r^{g}}
\label{g_pot}
\end{equation}
of a charge element of an electron at rest in $N^{g}$; here $\rho $ is the
charge density, $dV^{g}$ is a volume element of the charge and $r^{g}$ is
the distance from the charge to the observation point determined in $N^{g}$.

First, analogously to representing $r$ as $r=ct$ in an inertial reference
frame \cite[p. 416]{griffiths}, $r^{g}$ is expressed as $r^{g}=\bar{c}^{g}t$
in $N^{g}$. Assuming that ${\bf g}\cdot {\bf r/}2c^{2}$ $<<$ $1$ (weak
gravitational fields) we can write:
\begin{equation}
\left( r^{g}\right) ^{-1}\approx r^{-1}\left( 1-\frac{{\bf g}\cdot {\bf r}}{%
2c^{2}}\right) .  \label{r_g}
\end{equation}

\noindent Substituting $\left( r^{g}\right) ^{-1}$ in (\ref{g_pot}) gives
the potential (\ref{1fermi}) obtained by Fermi.

The second change in (\ref{g_pot}) is a Li\'{e}nard-Wiechert-like (or rather
anisotropic) volume element $dV^{g}$ (not coinciding with the actual volume
element $dV$) which arises in $N^{g}$ on account of the average anisotropic
velocity of light there. The origin of $dV^{g}$ is analogous to the origin
of the Li\'{e}nard-Wiechert volume element \cite[p. 418]{griffiths} $%
dV^{LW}=dV/\left( 1-{\bf v\cdot n/}c\right) $ of a charge moving at velocity
${\bf v}$ with respect to an inertial observer $I$, where ${\bf n=r/}r$ and $%
{\bf r}$ is the position vector at the retarded time. This can be explained
in terms of the ''information-collecting sphere'' of Panofsky and Phillips
\cite[p. 342]{panofski} used in the derivation of the Li\'{e}nard-Wiechert
potentials (similar concepts are employed by Griffiths \cite[p. 418]
{griffiths}, Feynman \cite[p. 21-10]{feynman}, and Schwartz \cite[p. 213]
{schwartz}). The Li\'{e}nard-Wiechert volume element $dV^{LW}$ of a charge
appears greater than $dV$ (in the direction of its velocity) due to its
greater contribution to the potential since it ''stays longer within the
information-collecting sphere'' \cite[p. 343]{panofski} sweeping over the
charge at the velocity of light $c$ in $I$. By the same argument the
anisotropic volume element $dV^{g}$ also appears different from $dV$ in $%
N^{g}$: in a direction opposite to ${\bf g}$ the velocity of the
information-collecting sphere (which propagates at the velocity of light (%
\ref{c_g}) in $N^{g}$) is smaller than $c$ since for light propagating
against~${\bf g}$ we have ${\bf g}\cdot {\bf r}=-gr$ in (\ref{c_g});
therefore an elementary volume $dV^{g}$ of the electron charge stays longer
within the sphere and contributes more to the potential in $N^{g}$.

Consider a charge of length $l$ at rest in $N^{g}$ placed along ${\bf g}$.
The time for which the information-collecting sphere sweeps over the charge
in $N^{g}$ is
\[
\Delta t^{g}=\frac{l}{\bar{c}^{g}}=\frac{l}{c\left( 1+{\bf g}\cdot {\bf r}%
/2c^{2}\right) }\approx \Delta t\left( 1-\frac{{\bf g}\cdot {\bf r}}{2c^{2}}%
\right) ,
\]
where $\Delta t=l/c$ is the time for which the information-collecting sphere
propagating at speed $c$ sweeps over an inertial charge of the same length $l
$ in its rest frame. When the information-collecting sphere moves against $%
{\bf g}$ in $N^{g}$ its velocity is smaller than $c$. Therefore the charge
stays longer within the sphere since $\Delta t^{g}>\Delta t$ (${\bf g}\cdot
{\bf r}=-gr$) and its contribution to the potential is greater. This is
equivalent to saying that the greater contribution comes from a charge of a
larger length $l^{g}$ which for the same time $\Delta t^{g}$ is swept over
by an information-collecting sphere propagating at velocity $c$:
\[
l^{g}=\Delta t^{g}c=l\left( 1-\frac{{\bf g}\cdot {\bf r}}{2c^{2}}\right) .
\]
The anisotropic volume element which corresponds to such an apparent length $%
l^{g}$ is obviously
\begin{equation}
dV^{g}=dV\left( 1-\frac{{\bf g}\cdot {\bf r}}{2c^{2}}\right) .  \label{dV_g}
\end{equation}
Substituting (\ref{r_g}) and (\ref{dV_g}) into (\ref{g_pot}) we obtain the
scalar potential of a charge element $\rho dV^{g}$ of the electron
\[
d\varphi ^{g}=\frac{1}{4\pi \epsilon _{0}}\frac{\rho dV^{g}}{r^{g}}=\frac{1}{%
4\pi \epsilon _{0}}\frac{\rho dV}{r}\left( 1-\frac{{\bf g}\cdot {\bf r}}{%
2c^{2}}\right) ^{2}
\]
or if we keep only the terms proportional to $c^{-2}$ we get
\begin{equation}
d\varphi ^{g}=\frac{\rho }{4\pi \epsilon _{0}r}\left( 1-\frac{{\bf g}\cdot
{\bf r}}{c^{2}}\right) dV.  \label{pot_g}
\end{equation}
As seen from (\ref{pot_g}) making use of $dV^{g}$ instead of $dV$ accounts
for the $1/2$ factor in (\ref{1fermi}). Now we can calculate the electric
field of a charge element $\rho dV^{g}$ of an electron at rest in $N^{g}$ by
using only the scalar potential (\ref{pot_g}):
\begin{equation}
d{\bf E}^{g}=-\nabla d\varphi ^{g}=\frac{1}{4\pi \epsilon _{o}}\left( \frac{%
{\bf n}}{r^{2}}-\frac{{\bf g\cdot n}}{c^{2}r}{\bf n}+\frac{{\bf g}}{c^{2}r}%
\right) \rho dV.  \label{d_E}
\end{equation}
\noindent The distortion of the electric field (\ref{d_E}) is caused by the
anisotropic velocity of light (\ref{c_g}) in $N^{g}$. For the distorted
field of the whole electron charge we find
\begin{equation}
{\bf E}^{g}=\frac{1}{4\pi \epsilon _{o}}\int \left( \frac{{\bf n}}{r^{2}}-%
\frac{{\bf g\cdot n}}{c^{2}r}{\bf n}+\frac{{\bf g}}{c^{2}r}\right) \rho dV{.}
\label{E_g}
\end{equation}
The self-force with which the field of an electron interacts with an element
$\rho dV_{1}^{g}$ of its charge is
\begin{equation}
d{\bf F}_{self}^{g}=\rho dV_{1}^{g}{\bf E}^{g}=\frac{1}{4\pi \epsilon _{o}}%
\int \left( \frac{{\bf n}}{r^{2}}-\frac{{\bf g}\cdot {\bf n}}{c^{2}r}{\bf n}+%
\frac{{\bf g}}{c^{2}r}\right) \rho ^{2}dVdV_{1}^{g}{.}  \label{dF_g}
\end{equation}
Due to the distorted electric field (\ref{E_g}) of an electron at rest in $%
N^{g}$ the mutual repulsion of its charge elements does not cancel out. As a
result a non-zero self-force with which the electron acts upon itself
arises:
\begin{equation}
{\bf F}_{self}^{g}=\frac{1}{4\pi \epsilon _{o}}\int \int \left( \frac{{\bf n}%
}{r^{2}}-\frac{{\bf g}\cdot {\bf n}}{c^{2}r}{\bf n}+\frac{{\bf g}}{c^{2}r}%
\right) \rho ^{2}dVdV_{1}^{g}{.}  \label{F_g}
\end{equation}
\noindent After taking into account the explicit form (\ref{dV_g}) of $%
dV_{1}^{g}$ (\ref{F_g}) becomes
\[
{\bf F}_{self}^{g}=\frac{1}{4\pi \epsilon _{o}}\int \int \left( \frac{{\bf n}%
}{r^{2}}-\frac{{\bf g}\cdot {\bf n}}{c^{2}r}{\bf n}+\frac{{\bf g}}{c^{2}r}%
\right) \left( 1-\frac{{\bf g}\cdot {\bf r}}{2c^{2}}\right) \rho ^{2}dVdV_{1}%
{.}
\]
Assuming a spherically symmetric distribution of the electron charge \cite
{lorentz} and following the standard procedure of calculating the self-force
\cite{podolsky} we get:
\begin{equation}
{\bf F}_{self}^{g}=\frac{U}{c^{2}}{\bf g},  \label{F_g1}
\end{equation}
where
\begin{equation}
U=\frac{1}{8\pi \epsilon _{o}}\int \int \frac{\rho ^{2}}{r}dVdV_{1}
\label{U}
\end{equation}
\noindent is the electrostatic energy of the electron. The famous $4/3$
factor does not appear in (\ref{F_g1}) since the correct volume element (\ref
{dV_g}) was used in (\ref{F_g}). As $U/c^{2}$ is the mass associated with
the energy $U$ of the electron field, (\ref{F_g1}) obtains the form:
\begin{equation}
{\bf F}_{self}^{g}=m^{g}{\bf g,}  \label{F=mg}
\end{equation}
\noindent where $m^{g}=U/c^{2}$ is interpreted as the electron passive
gravitational mass which is entirely electromagnetic in origin in the case
of the classical electron.

As (\ref{F=mg}) shows an electron whose worldline is not geodesic (since it
is at rest in $N^{g}$) is subjected to the self-force ${\bf F}_{self}^{g}$
which turns out to be equal to what is traditionally called the
gravitational force. The self-force ${\bf F}_{self}^{g}$ is electromagnetic
since it originates from the unbalanced repulsion of the electron charge
elements due to their distorted fields (\ref{d_E}) which in turn are caused
by the average anisotropic velocity of light in $N^{g}$. Thus the classical
electromagnetic mass theory in conjunction with the general-relativistic
corollary of the average anisotropic velocity of light in a gravitational
field provides an insight into the nature of the force acting on the
deviated from its geodesic path classical electron - the self-force ${\bf F}%
_{self}^{g}$ is inertial since it resists that deviation and is
electromagnetic in origin.

In general, if a charged classical particle is prevented from following a
geodesic path in a gravitational field, its electric field distorts and a
self-force resisting the deformation of the particle's field arises. That
force is inertial since it opposes the deviation of the particle from
maintaining its non-resistant (inertial) motion and is electromagnetic in
origin since it results from the unbalanced repulsion between the charged
elements of the particle.

Let us now see whether the approach outlined above predicts that a falling
classical electron is subjected to no force as required by general
relativity. To verify this let us calculate the electric field of an
electron falling in the Earth's gravitational field with an acceleration $%
{\bf a}={\bf g}$. We note that the Li\'{e}nard-Wiechert potentials must
include the correction due to the average anisotropic velocity of light in $%
N^{g}$:
\begin{equation}
\varphi ^{g}\left( r,t\right) =\frac{e}{4\pi \epsilon _{o}r}\frac{1}{1-{\bf %
v\cdot n}/c}\left( 1-\frac{{\bf g}\cdot {\bf r}}{c^{2}}\right)  \label{LWs_g}
\end{equation}
\begin{equation}
{\bf A}^{g}\left( r,t\right) =\frac{e}{4\pi \epsilon _{o}c^{2}r}\frac{{\bf v}%
}{1-{\bf v\cdot n}/c}\left( 1-\frac{{\bf g}\cdot {\bf r}}{c^{2}}\right) .
\label{LWv_g}
\end{equation}
The electric field of an electron falling in $N^{g}$ (and considered
instantaneously at rest in $N^{g}$ \cite{instant}) obtained from (\ref{LWs_g}%
) and (\ref{LWv_g}) is:
\[
{\bf E}=-\nabla \varphi ^{g}-\frac{\partial {\bf A}^{g}}{\partial t}=\frac{e%
}{4\pi \epsilon _{o}}\left[ \left( \frac{{\bf n}}{r^{2}}+\frac{{\bf g\cdot n}%
}{c^{2}r}{\bf n}-\frac{{\bf g}}{c^{2}r}\right) +\left( -\frac{{\bf g}\cdot
{\bf n}}{c^{2}r}{\bf n}+\frac{{\bf g}}{c^{2}r}\right) \right] ,
\]
\noindent which reduces to the Coulomb field \cite{rad}
\begin{equation}
{\bf E}=\frac{e}{4\pi \epsilon _{o}}\frac{{\bf n}}{r^{2}}.  \label{E_in}
\end{equation}
Therefore the instantaneous electric field of a falling electron is not
distorted. This demonstrates that the repulsion of its charge elements is
balanced, thus producing no self-force. This result sheds light on the
question why in general relativity an electron is falling in a gravitational
field ''by itself'' with no force acting on it. As (\ref{E_in}) shows, the
only way for an electron to compensate the anisotropy in the propagation of
light and to preserve the Coulomb shape of its electric field is to fall
with an acceleration ${\bf g}$. If the electron is prevented from falling
its electric field distorts (disturbing the balance of the repulsion of its
charge elements), the self-force (\ref{F=mg}) appears and tries to force the
electron to fall in order to eliminate the distortion of its field; if the
electron is left to fall its Coulomb field restores, the repulsion of its
charge elements cancels out and the self-force disappears \cite{geodesic}.

To summarize, the study of the open question in general relativity - what is
the nature of the force acting upon a particle deviated from its geodesic
path - by taking into account the classical electromagnetic mass theory
provides an insight not only into that question (in the case of the
classical electron) but also into the question why a falling electron (whose
worldline is geodesic) is subjected to no force.


\begin{thebibliography}{99}
\bibitem{classical}  We consider a classical electron since there exists no
quantum-mechanical description of the electron structure at this moment.
However, if a classical treatment succeeds in explaining the nature of the
force acting upon an electron whose world line is not geodesic, this will
imply that a corresponding quantum-mechanical description of that force is
also possible.
\bibitem{instant}  We consider the instantaneous electron field in order to
separate its Lorentz contraction from the distortion caused by the electron
acceleration and the anisotropic velocity of light.
\bibitem{rad}  It is clear from here that a falling electron does not
radiate since its electric field does not contain the radiation $r^{-1}$
terms \cite{petkov}. If those terms were present in the field of a falling
electron this would constitute a contradiction with the principle of
equivalence.
\bibitem{geodesic}  The behaviour of the classical electron in a
gravitational field as described by general relativity and the classical
electromagnetic mass theory can be summarized in the following way: the
worldline of an electron which preserves the shape of its Coulomb field is
geodesic and represents a free non-resistantly moving electron; if the field
of an electron is distorted, its worldline is not geodesic and the electron
is subjected to a self-force on account of its own distorted field.
\end{thebibliography}
\end{document}